\documentstyle[12pt]{aipproc}
\begin{document}\setlength{\unitlength}{1mm}
\title{Vacuum Condensates and the 
Anomalous Magnetic Moment of a Dirac Fermion}
\author{Victor Elias, Kevin B. Sprague, and Ying Xue$^{\dagger}$}
\address{$^{\dagger}$Department of Applied Mathematics\\
The University of Western Ontario\\
London, Ontario N6A 5B7 CANADA\\}
\maketitle
\begin{abstract}
We address anticipated fermion-antifermion and dimension-4 gauge-field vacuum-condensate
contributions to the magnetic portion of the fermion-photon vertex function in
the presence of a vacuum with nonperturbative content, such as that of QCD.  We
discuss how inclusion of such condensate contributions may lead to a {\it vanishing}
anomalous
magnetic moment, in which case vacuum condensates may account for the apparent
consistency between constituent quark masses characterizing baryon magnetic
moments and those characterizing baryon spectroscopy.
\end{abstract}
\renewcommand{\thesection}{\arabic{section}}
\renewcommand{\thesubsection}{}
\newpage
\section{INTRODUCTION} \label{introduction}
\renewcommand{\theequation}{1.\arabic{equation}}
\setcounter{equation}{0}

One of the earliest successes of quantum electrodynamics (QED) was the
successful calculation of the anomalous magnetic moment of an 
electron.$^{\cite{schwinger,feynman}}$  Such a calculation is, of course, performed
within the context of a purely perturbative quantum field theory whose vacuum
does not permit the formation of such nonperturbative exotica as
electron-positron condensates and multiple-photon condensates.  This is in
contrast to the vacuum characterizing quantum chromodynamics (QCD), which
clearly has nonperturbative content, as evident from the occurrence of
quark-antiquark, multiple-gluon, and higher dimensional quark-gluon QCD-vacuum
condensates which characterize QCD sum-rule applications.$^{\cite{shifman}}$
Although QCD has had a great deal of success in predicting hadronic physics
(particularly at high energies), it almost never is made to address the
low-energy properties of its fundamental fermions ({\it i.e.}, quarks), even
though analogous lepton properties are convincingly addressed by QED.

Part of the reason for this, of course, is confinement.  We are completely
incapable of measuring the magnetic moment of a quark in the same way that we
can measure the quantity $(g-2)$ for a lepton.  The strong coupling is itself ambiguous at low
momenta.  However, there does exist model-dependent but well-established and
simple phenomenology that successfully determines the magnetic moments of
baryons in terms of the magnetic moments of their constituent
quarks.$^{\cite{beg,ruj}}$  The quark masses characterizing quark-magnetons within
this phenomenology are suprisingly consistent with those characterizing hadron
spectroscopy within the nonrelativistic quark
model.$^{\cite{ruj,isgur,perkins}}$ 
Thus, we do appear to have an indirect way of talking about quark magnetic
moments, at least within the context of an hadronic environment, suggesting that
there may be a value in having QCD address the QED-like question of what a
quark's anomalous magnetic moment really is.

In Section \ref{review}, we review how the anomalous magnetic moment of a Dirac
fermion is obtained in QED.  The result ${\cal K} F_2(0)=e^2Q^2/8\pi^2$ ($Q$ is
the fermion charge) is obtained {\it entirely} from the single-photon-exchange
vertex correction of Fig. 1.$^{\cite{feynman}}$  If such a fermion were a quark, it
would also couple to massless QCD gluons.  An inevitable QCD modification of the
perturbative calculation would entail the replacement of the momentum-$k$ photon
in Fig. 1 with gluons, including appropriate colour factors $\lambda^a_{ij}/2$ at
each terminus of the gluon line $G^a$.  The net effect of incorporating such
gluon lines would be to augment the factor $e^2Q^2$ in ${\cal K} F_2(0)$ with an
additional factor of $4g_s^2/3$ [$4/3$ is the $SU(3)_c$ group theoretical factor
$\lambda^a_{ij} \lambda^a_{ji}/4$].  Since we can anticipate that $\alpha_s >>
\alpha$ at low momenta, we would (naively) anticipate a quark anomalous magnetic
moment approximately given by ${\cal K} F_2(0) \cong 2 \alpha_s(0)/3 \pi$.  Such
an anomalous moment is likely to be sufficiently large to compromise the success
of the static quark model in calculating baryon magnetic 
moments.$^{\cite{beg,ruj}}$

An immediate problem arises, of course, from attempting at all to utilize
$\alpha_s$ in the soft-momentum (large-distance) limit.  Mattingly and Stevenson
have argued$^{\cite{mat}}$ that $\alpha_s$ in this limit should approach an
infrared-stable fixed point somewhat smaller than unity, a result corroborated
by recent work$^{\cite{babou}}$ exploring the possible linkage between
linear-sigma-model hadronic phenomenology and low-energy QCD.  It has also been
argued$^{\cite{fomin}}$ that values of $\alpha_s$ near unity ($\alpha_s = 3/\pi$)
induce chiral-symmetry breaking responsible for the low-energy transition from
QCD to a chiral-lagrangian theory of mesons.  Despite the arguments of ref.
\cite{mat}, there is contradictory evidence that an infrared-stable fixed point
cannot occur unless $n_f$ is substantially larger than
three.$^{\cite{gardi,elias98,chish}}$   Arguments have also been advanced for the
existence of an order-unity infrared attractor for the $n_f=3$ QCD
$\beta$-function that effectively corresponds to the (finite) coupling strength
for the infrared boundary of QCD.  The idea of introducing $\alpha_s(0)$ as a
phenomenological parameter has support from other empirical contexts, as
well.$^{\cite{dok}}$

In any case, we see that there exists motivation for having the infrared limit
of QCD characterized by an effective value of $\alpha_s$ near
unity.\footnote{Refs \cite{mat} and \cite{babou} respectively advance values of
$0.82$ and $0.72$ for $\alpha_s$ in the infrared region.  Ref. \cite{chish}
suggests an infrared attractor at $\alpha_s/\pi \cong 0.4$.  This latter
estimate is based upon $[2|2]$ and $[1|3]$ Pad\'e approximants for the
higher-than-one-loop terms in the $n_f=3$ $\beta$-function, in conjunction with
a large and negative asymptotic-error-formula estimate of that
$\beta$-function's five loop term; {\it i.e.,} note that the value of $x_d
(= \alpha_s/ \pi)$ in Figs. 10 and 11 of ref. \cite{chish} is near $0.4$ 
when $R_4 (\equiv \beta_4 / \beta_0)$ 
is equal to $-850$, as predicted in ref. \cite{elias98}.}
This suggests a large QCD contribution to the quark's anomalous magnetic
moment, ${\cal K} F_2(0) \cong 2\alpha_s(0)/3\pi \cong 0.2$, arising {\it
entirely} from purely perturbative single-gluon exchange.  Such a $20\%$ increase
in the quark's magnetic moment has significant consequences when applied to
constituent-quark-model estimates of nucleon (and other baryon) magnetic
moments.  The magneton characterizing such estimates is consistent with
baryon constituent quark masses $m=m_d \cong 360 \;MeV$.$^{\cite{perkins}}$  Given empirical
constraints, however, a $20\%$ increase in the quark's true  magnetic moment
necessarily entails a concomitant $20\%$ increase in the true constituent quark
mass ($m_{qk}$) appearing in the quark magneton:
\begin{equation} \label{muEmp}
\mu_{empirical}=e^2/(2 \cdot 360\; MeV/c^2) = (e^2/2m_{qk})[1+{\cal K} F_2(0)]~.
\end{equation}
Based on the above estimate, Eq. (\ref{muEmp}) suggests that the true constituent
quark mass is elevated to $430\;MeV/c^2$, a value significantly larger than the
constituent light quark mass characterizing hadron spectroscopy.  Of course,
there is latitude in this estimate (depending on $\alpha_s(0)$), but the net
result is clearly a disequilibration of spectroscopic and magnetic-moment
constituent-quark masses.

The above arguments suggest the need for a more careful attempt to ascertain the
full QCD contribution to the anomalous magnetic moment of its
fundamental-fermion constituents.  Such an attempt must necessarily include
incorporation of QCD-vacuum condensates which have already been
introduced into field-theoretical calculations of QCD vacuum-polarization
diagrams and the two-current correlation functions associated with QCD sum-rule
applications.$^{\cite{shifman,pascual}}$  The same field-theoretical techniques that
enable the evaluation of vacuum condensate contributions to two-point functions
can be utilized to determine corresponding vacuum condensate contributions to
QCD three-point functions as well.

In Section \ref{fermContrib}, we show how fermion-antifermion condensate
contributions to the electromagnetic vertex function can be calculated, using
sum-rule methods delineated in ref. {\cite{bagan94}.  For simplicity, we work
within QED, but assume the presence of a chiral noninvariant vacuum 
capable of forming such a
condensate.  The startling result of this calculation is that the real part of
the magnetic portion of the vertex correction vanishes over the soft momentum
range $0 < q^2 < 4m_{qk}^2$, a result suggesting the absence of {\it any}
fermion-antifermion condensate contribution to ${\cal K} F_2(0)$.  However, this
same magnetic term exhibits an imaginary part over the same kinematic region, in
apparent correspondence (within a QCD context) to a dispersive contribution {\it
below} the threshold for a "physical" quark-antiquark intermediate state.

In Section \ref{gluContrib}, we calculate the covariant-gauge dimension-4 
gluon-condensate contributions to the electromagnetic vertex function.  The use of a
covariant gauge is shown to be necessary in order to retain the
configuration-space translational invariance within the Feynman amplitude that
is required for the factorization of external lines from the truncated vertex
Green's function.  Both the magnetic and the electric-charge contribution to the
vertex correction are shown to be highly divergent on-shell.  The naive
interpretation of this result is to say that quarks exhibit an infinite magnetic
moment, since the divergence of the magnetic contribution cannot be removed by
renormalization, which has already been applied to ensure that the electric charge
form-factor is unity at $q^2=0$.  Obviously, particles with infinite magnetic
moments can never appear to be free, consistent with confinement expectations
(at least for quarks).  However, a true demonstration of confinement on this
basis would have to account for the full cancellation of such divergent
magnetic-moment contributions within colour-singlet (only) constructions of
quarks and gluons.

In Section \ref{discussion}, we attempt to interpret the salient features of
the previous two sections.  We argue that the kinematic extension of the
dispersive domain found in Section \ref{fermContrib} can be explained as a
consequence of the Goldstone theorem -- that the underlying assumptions of the
calculation necessarily entail the existence of zero-mass Goldstone bosons as
part of the physical particle spectrum.  We then discuss the divergent 
gluon-condensate contributions to the electromagnetic vertex, and argue that such
contributions must absorb two powers of the coupling to be renormalization-group
invariant.  This result suggests that these effects are, strictly speaking, a
tree-order contribution, as opposed to a one-loop order contribution analogous
to the purely perturbative case delineated in Section \ref{review}.  If one then
reformulates the perturbative series appropriately, the anomalous magnetic
moment is seen to vanish entirely as a consequence of the differing degrees of
divergence between magnetic and electric-charge contributions to the 
gluon-condensate component of the vertex correction.  Such an interpretation, if
correct, would account for the compatibility of the naive magnetic moment of
constituent quarks with the constituent-quark masses that characterize hadron
spectroscopy.

\newpage
\section{Methodology of the Purely Perturbative Case} \label{review}
\renewcommand{\theequation}{2.\arabic{equation}}
\setcounter{equation}{0}

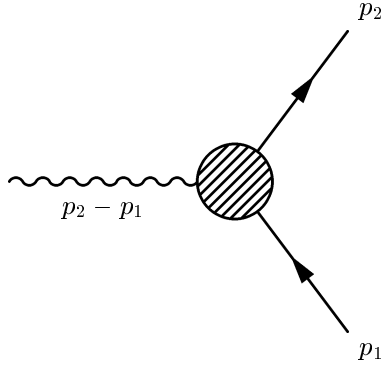
Purely perturbative contributions to the fermion-antifermion-photon Green
function $G^\mu$, as depicted in Fig. 2, can be expressed in terms of 
time-ordered products of Heisenberg fields $\psi^h$, $\bar{\psi}^h$ and $A_\mu^h$:
\begin{equation} \label{Gheis}
G_\mu(p_2,p_1) = \int d^4x' \int d^4y'
<0|T\psi^h (x') A^h_\mu(0)\bar{\psi}^h (y')|0> 
 e^{i p_2 \cdot x'}e^{-i p_1 \cdot y'},
\end{equation}
To one-loop order, the correction to this vertex is obtained in terms of
interaction-picture fields $\psi$, $\bar{\psi}$, $A_\mu$ via the Wick-Dyson
expansion of the time-ordered product within (\ref{Gheis}),
\begin{eqnarray} \label{T}
&&<0|T\psi^h (x') A^h_\mu(0)\bar{\psi}^h (y')|0> \nonumber\\
&&~~~=
<0|T\psi(x')\; exp \left[-ieQ\int d^4w \; \bar{\psi}(w) \gamma^\sigma \psi(w) 
A_\sigma(w) \right] A_\mu(0)\bar{\psi} (y')|0> ~,
\end{eqnarray}
in which case
\begin{eqnarray}\label{Gexpansion}
&&G_\mu(p_2,p_1) = \int d^4x' d^4y' e^{i p_2 \cdot x'} e^{-i p_1 \cdot
y'}\nonumber\\ 
&&~~~~~~~~~~~~~~~~\left\{
(-ieQ) \int d^4 w <0|T \psi (x') \bar{\psi}(w)|0> \gamma^\sigma
<0|T \psi (w) \bar{\psi}(y')|0> \right.\nonumber\\
&&~~~~~~~~~~~~~~~~~~~~~~~~~~~~~~\cdot <0|T A_\sigma(w) A_\mu (0)|0> \nonumber\\
&&~~~~~~~~~~~~~~~~ + (-ieQ)^3 \int d^4 x \;d^4y \; d^4z <0|T \psi (x') \bar{\psi}(x)|0>
\gamma^\tau <0|T\psi(x) \bar{\psi} (y)|0> \nonumber\\
&&~~~~~~~~~~~~~~~~~~~~~~~~~~~~~~~~\cdot \gamma^\sigma
<0|T\psi(y) \bar{\psi} (z)|0> \gamma^\rho <0|T\psi(z) \bar{\psi} (y')|0> 
\nonumber\\
&&~~~~~~~~~~~~~~~~~~~~~~~~~~~~~~~~
\cdot <0|T A_\tau (x) A_\rho (z)|0><0| T A_\sigma (y) A_\mu (0)|0> \nonumber\\
&&~~~~~~~~~~~~~~~~\left. + {\cal O}[(ieQ)^5]\right\} 
\end{eqnarray}
One obtains (\ref{Gexpansion}) by dropping all normal-ordered terms from the
Wick-Dyson expansion of the time-ordered product on the right side of (\ref{T}),
as normal-ordering necessarily places annihilation operators strictly against
the purely perturbative vacuum ket $|0>$:
\begin{equation} \label{pertNproduct}
<0|:(Product~of~any~number~of~field~operators):|0>=0
\end{equation}
Eq. (\ref{pertNproduct}) is no longer automatic if the vacuum is assumed to have
nonperturbative content, notationally delineated by replacing $|0>$ with the
nonperturbative ({\it i.e.,} QCD) vacuum $|\Omega>$. 
Indeed, such nonperturbative vacuum expectation values provide the explicit 
mechanism by which QCD-vacuum condensates contribute to field-theoretical
current-correlation functions responsible for QCD sum-rules.$^{\cite{pascual}}$
This same mechanism enables vacuum condensates to contribute to Feynman
amplitudes in general.  

Before we discuss such nonperturbative contributions to the
fermion-antifermion-photon vertex of Fig. 2, it will prove methodologically
useful for us to outline how the purely perturbative contribution is obtained.
The two-field vacuum expectation values in (\ref{Gexpansion}) are simply
configuration-space Feynman propagators,
\begin{equation}
\label{fermionProp}
<0|T\psi (x) \bar{\psi}(y)|0> 
= i \int \frac{d^4q}{(2 \pi)^4} e^{-i q \cdot (x-y)}
\frac{(\gamma \cdot q) +m}{q^2 - m^2+i|\epsilon|},
\end{equation}
\begin{eqnarray} \label{photonProp}
<0|TA_{\tau} (x) A_{\rho} (z)|0>
= -i\int \frac{d^4k}{(2 \pi)^4} 
 e^{-i k \cdot (x-z)}\left(\frac{g_{\tau
 \rho}}{k^2+i|\epsilon|}-\frac{(1-\xi)k_\tau k_\rho}{k^4}\right),
\end{eqnarray}
where $\xi$ is an (irrelevant) gauge-parameter coefficient of the longitudinal
propagator component.  The translational invariance of (\ref{fermionProp}) and
(\ref{photonProp}) enables factorization of external lines from the full
one-loop contribution to $G_\mu$:
\renewcommand{\theequation} {2.7\alph{equation}}
\setcounter{equation}{0}
\begin{eqnarray} \label{qedGreen}
G_\mu(p_2,p_1) &=& i\left(\frac{\not p_2 + m}{{p_2}^2 - m^2}\right)
\left(\frac{-ig_{\mu\sigma}}{(p_1-p_2)^2}\right)
[(-ieQ)\{\gamma^\sigma+\Gamma^\sigma(p_2,p_1) \}]
\nonumber\\&~& \cdot i\left(\frac{\not {p_1} + m}{{p_1}^2 - m^2}\right)~,
\end{eqnarray}
\begin{eqnarray} \label{qedVertexInt}
&& \Gamma^\sigma(p_2,p_1) \nonumber\\ &&~~
= -\frac{ie^2Q^2}{(2\pi)^4} \int \frac{d^4k}{k^2} \gamma^\tau
\left(\frac{\not {p_2-k} + m}{{p_2-k}^2 - m^2}\right) \gamma^\sigma
\left(\frac{\not {p_1-k} + m}{{p_1-k}^2 - m^2}\right) \gamma_\tau
 + {\cal O}(e^4)
\end{eqnarray}
\renewcommand{\theequation} {2.\arabic{equation}}
\setcounter{equation}{7}
The truncated vertex function (\ref{qedVertexInt}) corresponds to the Feynman
diagram in Fig. 1.  The leading contribution to the anomalous magnetic moment of
the fermion can be extracted from an explicitly finite coefficient $S(q^2)$
within this (unrenormalized) vertex function;
\begin{equation}\label{qedVertexStructure}
\Gamma^\sigma (p_2,p_1) \equiv e^2Q^2 \left[ R((p_2-p_1)^2) \gamma^\sigma +
\frac{2 S((p_2-p_1)^2)}{m}(p_1^\sigma+p_2^\sigma) \right]
\end{equation}
where $S(0)$, as extracted from explicit evaluation of (\ref{qedVertexInt}) is
found to be$^{\cite{feynman}}$
\begin{equation}\label{S0_feyn}
S(0)=-\frac{1}{32} \pi^2
\end{equation}
Making use of the Gordon decomposition relation
\begin{equation}
\bar{u}(p_2)(p_1^\sigma + p_2^\sigma)u(p_1) =
\bar{u}(p_2)[2m\gamma^\sigma - i \sigma^{\sigma \mu}(p_2 - p_1)_\mu]u(p_1)
\end{equation}
one finds that the unrenormalized vertex function may be expressed in terms of
the momentum transfer $q \equiv (p_2-p_1)$:
\begin{eqnarray} \label{211}
&&\bar{u}(p_2)[\gamma^\tau +\Gamma^\sigma(p_2,p_1)]u(p_1) \nonumber\\
&&~~=\bar{u}(p_2) \left[(1+e^2Q^2[R(q^2)+4S(q^2)])\gamma^\tau - \frac{2ie^2Q^2}{m}
S(q^2)\sigma^{\tau \mu} q_\mu \right]u(p_1) \nonumber\\
&&~~\equiv Z \bar{u}(p_2) \left[F_1(q^2)\gamma^\tau + \frac{i}{2m}\sigma^{\tau \mu}
q_\mu{\cal K} F_2(q^2) \right]u(p_1)
\end{eqnarray}
where $F_1(q^2)$ is the fermion's charge form-factor and ${\cal K} F_2(0)$  is the
fermion's anomalous magnetic moment.
The requirement that $F_1(0)=1$ imposes the following rescaling of the
amplitude: 
\begin{equation}
Z=1+e^2Q^2[(R(0)+4S(0)]+{\cal O}(e^4)
\end{equation}
in which case the anomalous magnetic moment is given by
\begin{eqnarray}\label{amm}
{\cal K} F_2(0) &=& \frac{-4e^2Q^2 S(0)}{1+e^2Q^2[R(0)+4S(0)]}\nonumber\\
&=& -4e^2Q^2S(0)+{\cal O}(e^4)
\end{eqnarray}
The finite calculated value of $S(0)$ [eq.(\ref{S0_feyn})] ensures a finite 
anomalous magnetic moment:$^{\cite{schwinger,feynman}}$
\begin{equation}
{\cal K} F_2(0) = \frac{\alpha}{2 \pi} + {\cal O}(\alpha^2)
\end{equation}
By contrast, the coefficient $R(q^2)$ within (\ref{qedVertexStructure}) is
divergent at $q^2=0$.  Of course, additional self-energy and bremmstrahlung diagrams also
contribute to the full vertex contribution, and these contributions are necessary
to obtain a finite (and correct) answer for the renormalized charge-form-factor
slope $F_1'(0)$.  However, such contributions only enter the evaluation of
$R(q^2)$ within the full unrenormalized vertex; $S(q^2)$ [and, consequently,
${\cal K} F_2(0)$] arises {\it entirely} from the Feynman amplitude 
(\ref{qedVertexInt}).  In the two sections that follow we will respectively
calculate the fermion-antifermion condensate and gluon-condensate contributions
to $S(q^2)$, since $S(0)$ alone appears to determine the anomalous magnetic
moment.

\section{Fermion-Antifermion Condensate Contribution to the Electromagnetic
Vertex} \label{fermContrib}
\renewcommand{\theequation}{3.\arabic{equation}}
\setcounter{equation}{0}

The fermion-antifermion condensate contribution to the electromagnetic vertex
of Fig. 2 has been evaluated at length.$^{\cite{elias98-2}}$  We recapitulate the
essential features of this calculation here.  The Wick-Dyson expansion
for the time-ordered product of fermion-antifermion fields is
\begin{equation}\label{Torder}
T\psi(x) \bar{\psi}(y) = <0|T \psi(x) \bar{\psi} (y)|0> + :\psi(x)
\bar{\psi}(y):
\end{equation}
If the true vacuum $|\Omega>$ does not exhibit chiral-invariance, the
vacuum expectation value of (\ref{Torder}) will acquire not only the
perturbative configuration-space fermion propagator (\ref{fermionProp}) but also
a {\it nonperturbative propagator} component$^{\cite{bagan94,ynd}}$
\renewcommand{\theequation} {3.2\alph{equation}}
\setcounter{equation}{0}
\begin{equation}\label{Torder2}
<\Omega|T\psi(x) \bar{\psi}(y)|\Omega>=<0|T \psi(x) \bar{\psi} (y)|0> + 
<\Omega|:\psi(x) \bar{\psi}(y):|\Omega>
\end{equation}
\begin{eqnarray} \label{F}
<\Omega|:\psi(x) \bar{\psi}(y):|\Omega> &=& \int d^4k \; e^{-ik \cdot
(x-y)}(\not{k}+m) {\cal F}(k); \nonumber\\
\int d^4k \; {\cal F}(k)e^{-ik \cdot x} &\equiv& <f \bar{f}> 
\frac{J_1(m \sqrt{x^2})}{6m^2 \sqrt{x^2}}
\end{eqnarray}
\renewcommand{\theequation} {3.\arabic{equation}}
\setcounter{equation}{2}
In (3.2) $<f \bar{f}>$ is proportional (up to group theoretical factors) to the
fermion-antifermion condensate; $<f \bar{f}>=-<\bar{q} q>$ for the case of
$SU(3)_c$ fundamental-representation fermions - {\it i.e.,} quarks.  The
additional contribution to (\ref{Torder2}) enters the Feynman amplitude for the
vertex correction (as indicated symbolically in Fig. 3) by replacing
perturbative fermion propagators within (\ref{Gexpansion}) with the appropriate
nonperturbative propagator functions.  This net extra fermion-antifermion
contribution is given by
\begin{eqnarray}\label{GexpansionNP}
&&\Delta G_\mu(p_2,p_1) \nonumber\\ &&~
= \int d^4x' d^4y' e^{i p_2 \cdot x'} e^{-i p_1 \cdot
y'}
\left\{(-ieQ)^3 \int d^4 x \; d^4y \; d^4z 
<0|T \psi (x') \bar{\psi}(x)|0> \gamma^\tau \right.
\nonumber\\ &&~~~~~~\cdot
\left[<0|T\psi(x) \bar{\psi} (y)|0> 
\gamma^\sigma <\Omega|:\psi(y) \bar{\psi} (z):|\Omega> \right.
\nonumber\\ &&~~~~~~~~~~~~ \left. 
+ <\Omega|:\psi(x) \bar{\psi} (y):|\Omega>\gamma^\sigma 
<0|T\psi(y) \bar{\psi} (z)|0> \right] \gamma^\rho 
\nonumber\\ &&~~~~~~\cdot \left. 
<0|T\psi(z) \bar{\psi} (y')|0> 
<0| T A_\tau (x) A_\rho (z)|0><0|T A_\sigma (y) A_\mu (0)|0> \right\} 
\end{eqnarray}
[The term in which two nonperturbative vacuum expectation values appear
vanishes.$^{\cite{elias98-2}}$]

The momentum-space realization of (\ref{GexpansionNP}) is straightforward to
obtain by use of (\ref{F}) in conjunction with (\ref{fermionProp}) and
(\ref{photonProp}).  The explicit translational invariance of (\ref{F}) permits
the same factorization of external lines from the truncated vertex as occurs in
the purely perturbative case.  The contribution of Fig. 3a, for example, is
obtained from (\ref{qedVertexInt}) by replacing $[(k-p_2)^2-m^2]^{-1}$ with
$-i(2\pi)^4 {\cal F}(k-p_2)$; similarly the contribution of Fig. 3b is obtained by
replacing $[(k-p_1)^2-m^2]^{-1}$ with $-i(2\pi)^4 {\cal F}(k-p_1)$.  The net
effect of these changes is the following $<f \bar{f}>$-sensitive correction to
the truncated electromagnetic vertex:$^{\cite{elias98-2}}$:
\begin{eqnarray}
\Delta \Gamma^\mu(p_2,p_1) = -ie^2Q^2 &\{&~[-2 \not{p_1} \gamma^\mu
\not{p_2} + 4m(p_1^\mu+p_2^\mu)-2m^2\gamma^\mu] \;I \nonumber\\&~&
+ [2\gamma^\rho \gamma^\mu \not{p_2}+2  \not{p_1} \gamma^\mu \gamma^\rho -
8mg^{\mu \rho}] \; I_\rho - 2 \gamma^\rho \gamma^\mu \gamma^\sigma \; 
I_{\rho \sigma}\},
\end{eqnarray}
where the integrals $I, I_\rho, I_{\rho \sigma}$ (after trivial shifts in the
integration variable) are
\begin{eqnarray}
[I; \; I_\rho; \; I_{\rho \sigma}] &=& -i \int d^4k \frac{{\cal F}(k)[1; \;
k_\rho+p_{1 \rho}; \; (k_\rho+p_{1 \rho})(k_\sigma+p_{1
\sigma})]}{(k+p_1)^2[(k+p_1-p_2)^2-m^2]}\nonumber\\&~& 
-i\int d^4k \frac{{\cal F}(k)[1; \;
k_\rho+p_{2 \rho}; \; (k_\rho+p_{2 \rho})(k_\sigma+p_{2
\sigma})]}{(k+p_2)^2[(k+p_2-p_1)^2-m^2]}
\end{eqnarray}
The aggregate contribution to $S(q^2)$ is the coefficient responsible for the
anomalous magnetic moment within the vertex correction
(\ref{qedVertexStructure}), and is found to be$^{\cite{elias98-2}}$
\begin{eqnarray} \label{S}
\Delta S(q^2) = \frac{m^2}{3} \int_0^1 &&dz \frac{1-z}{[m^2(1-z)^2+q^2z]^2}
\left[ \frac{<f \bar{f}>(1-z)}{6m} \right. \nonumber\\ &&~\nonumber\\ &&
  - [5m^2(1-z)^2+q^2z(1-4z)] \; R_2(p_1z-p_2,mz) \nonumber\\&&~ \nonumber\\ 
&& \left.
  - [m^2q^2z(1-z)(3+5z)+q^4z^2(1+2z)] \; R_3(p_1z-p_2,mz) \right]
\end{eqnarray}
where for $p^2>0$
\begin{eqnarray}\label{A:R2}
& & R_2 (p, \; \mu) \equiv \int \frac{d^4 k \; {\cal F}(k)}{(p-k)^2 -
\mu^2 + i |\epsilon|} = \frac{<f \bar{f}>}{24m^3 p^2} ~~~~~~~~~~~~~~~~~~
~~~~ \nonumber\\ 
& &~~~~~~\times \left[ p^2 + m^2 - \mu^2 -
\sqrt{[p^2 -(m-\mu)^2][p^2 - (m+\mu)^2]} \right], 
\end{eqnarray}
\begin{eqnarray}\label{A:R3}
R_3 (p, \; \mu) & \equiv & \int \frac{d^4 k \; {\cal F}(k)}{\left[ (p-k)^2 
-
\mu^2 + i |\epsilon| \right]^2} \nonumber \\ 
& = & -\frac{<f\bar{f}>}{24m^3 p^2} \left[ 1 - \frac{p^2 + m^2 - 
\mu^2}{\sqrt{[p^2 
- (m - \mu)^2][p^2 - (m+\mu)^2]}} \right] .
\end{eqnarray} 
Explicit evaluation of the integral (\ref{S}) on the $p_1^2=p_2^2=m^2$ mass shell
in the region $0<q^2<4m^2$ yields the following remarkable result:
\newpage
\begin{eqnarray} \label{reS}
Re[\Delta S(q^2)]&=&  
\frac{m^2<f \bar{f}>}{3} \int_0^1 dz \frac{1-z}{[m^2(1-z)^2+q^2z]^2} 
\nonumber\\ &~&~~~
\cdot \left\{
\frac{1-z}{6m} -\frac{[5m^2(1-z)^2+q^2z(1-4z)][2m^2(1-z)+q^2z]}{24m^3[m^2(1-z)^2+q^2z]}
\right. \nonumber\\ &~&~~~~~~ \left.
+\frac{[m^2q^2z(1-z)(3+5z)+q^4z^2(1+2z)]}{24m^3[m^2(1-z)^2+q^2z]} \right\}
\nonumber\\ &=&0~~~~(0<q^2<4m^2)
\end{eqnarray}
If $0<q^2<4m^2$, the imaginary part of (\ref{S}) is nonzero, despite the fact
that the vertex momentum is {\it below} the kinematic threshold for the
production of a quark-antiquark pair:
\begin{eqnarray}\label{imS}
Im[\Delta S(q^2)] &=& \frac{<f \bar{f}>}{72m} \int_0^1 dz \frac{1-z}
{[m^2(1-z)^2+q^2z]^3} \nonumber\\ &~&~~\cdot \left\{
[5m^2(1-z)^2+q^2z(1-4z)]z \sqrt{q^2(4m^2-q^2)} \right. \nonumber\\ &~&~~~~~~
\left.
+[m^2q^2z(1-z)(3+5z)+q^4z^2(1+2z)] \frac{2m^2(1-z)+q^2z}{z \sqrt{q^2(4m^2-q^2)}}
\right \} \nonumber\\
&=& \frac{<f \bar{f}>}{12 m \sqrt{4m^2q^2-q^4}}.
\end{eqnarray}
The interpretation of these results will be deferred to Section
\ref{discussion}.  It should be noted here, however, that in the context of QCD,
$e^2<f \bar{f}> \rightarrow -g_s^2<q \bar{q}> \sim m^3$, where $m$ is understood
to be the gauge invariant dynamical mass arising from the chiral noninvariance
of the QCD vacuum.$^{\cite{pol}}$  Consequently, a single scale parameter $m$
characterizes $g_s^2 \Delta S(q^2)$.

\newpage
\section{The Gluon-Condensate Contribution to the Electromagnetic Vertex}
\label{gluContrib}
\renewcommand{\theequation}{4.\arabic{equation}}
\setcounter{equation}{0}

The Wick-Dyson expansion for a time-ordered product of gluon fields includes
both a configuration-space propagator and a normal-ordered piece:
\begin{equation} \label{Tscalars}
TA_\tau(x) A_\rho(z) = <0|T A_\tau (x) A_\rho (z)|0> + :A_\tau(x) A_\rho(z):
\end{equation}
If the QCD vacuum $|\Omega>$ has nonperturbative content, the vacuum expectation
value of (\ref{Tscalars}) will acquire an additional nonperturbative component
\begin{equation}\label{Tscalars2}
<\Omega|TA_\tau(x) A_\rho(z)|\Omega>=<0|T A_\tau(x) A_\rho(z)|0>+
<\Omega|:A_\tau(x) A_\rho(z):|\Omega>
\end{equation}
The first term on the right hand side of (\ref{Tscalars2}) is the
purely perturbative configuration-space propagator given by (\ref{photonProp}). 
When evaluated in a translationally-invariant ({\it i.e.,} covariant) gauge, as
required for the factorization of external lines from the truncated vertex
function, the second term of (\ref{Tscalars2}) is given by$^{\cite{bagan89}}$
\begin{equation}       \label{gluonCondProp}                                                                                                                                           
<\Omega|:A_\tau(x)A_\rho(z):|\Omega> 
= \frac{1}{8} \left[C(x-z)_\tau (x-z)_\rho
+ Eg_{\tau \rho}(x-z)^2\right]
\end{equation}
where the constants $C$ and $E$ are proportional to the dimension-4 
gluon-condensate $<G^2>$:
\renewcommand{\theequation}{4.4\alph{equation}}
\setcounter{equation}{0}
\begin{equation} \label{C}
C = \frac{<G^2>}{144} - \frac{<\Omega|:(\partial \cdot A)^2:|\Omega>}{24}
\end{equation}
\begin{equation} \label{E} 
E = -\frac{5<G^2>}{288} - \frac{<\Omega|:(\partial \cdot A)^2:|\Omega>}{48}
\end{equation}
\renewcommand{\theequation}{4.\arabic{equation}}
\setcounter{equation}{4}
and the gluon-condensate $<G^2>$ is defined to be
\begin{equation}
<G^2> =<0|:G^a_{\mu\nu}(0) G_{\mu\nu}^a(0):|0>
\end{equation}
with $G^a_{\mu \nu}$ the QCD field-strength tensor.
The $<(\partial \cdot A)^2>$ condensate is a reflection of the gauge dependence
of (\ref{gluonCondProp}); the cancellation of this condensate from several
gauge-invariant amplitudes has been demonstrated in ref. \cite{bagan94}.

The gluon-condensate contribution to the electromagnetic vertex is schematically
represented by Fig. 4, and corresponds to replacing $<0|T A_\tau (x) A_\rho
(z)|0>$ within (\ref{Gexpansion}) with the nonperturbative vacuum expectation
value (\ref{gluonCondProp}).  Upon inclusion of $SU(3)_c$ vertex structure, the
corresponding correction to the (untruncated) vertex Green's function is found
to be
\begin{eqnarray} \label{qcdGreen}
&& \Delta G^{\mu}(p_2,p_1) \nonumber\\&&~~~= (-ieQ)(-ig_s)^2
Tr\left(\frac{\lambda^a}{2}\frac{\lambda^b}{2} \right) 
\int d^4x'd^4y'd^4x \; d^4y \; d^4z \;
e^{i p_2 \cdot y'}e^{-i p_1 \cdot x'}  \nonumber\\
&&~~~~~~ \cdot \left[i \int \frac{d^4q_1}{(2\pi)^4} 
\frac{\not q_1 + m}{{q_1}^2 - m^2}e^{-iq_1 \cdot (y'-z)}\right] \gamma_\tau 
\left[i \int \frac{d^4q_2}{(2\pi)^4} \frac{\not q_2 + m}{{q_2}^2 - m^2}
e^{-iq_2 \cdot (z-y)}\right]\gamma_\sigma \nonumber\\
&&~~~~~~\cdot\left[i \int \frac{d^4q_2}{(2\pi)^4} 
\frac{\not q_3 + m}{{q_3}^2 - m^2}e^{-iq_3 \cdot (y-x)}\right]\gamma_\rho
\left[i \int \frac{d^4q_4}{(2\pi)^4} \frac{\not q_4 + m}{{q_4}^2 - m^2}
e^{-iq_4 \cdot (x-x')}\right] \nonumber\\
&&~~~~~~\cdot
\frac{1}{8} \left[C(x-z)^\rho(x-z)^\tau
+ g^{\rho\tau}E(x-z)^2\right] 
 \left[(-i {g}^{\mu \sigma}) \int \frac{d^4k'}{(2\pi)^4}
\frac{1}{k'^2}e^{ik' \cdot y}\right] \nonumber\\
&&~~~= \frac{i^2}{8}(-ieQ)(-ig_s)^2\left(\frac{1}{3}\sum_a 
\frac{\delta^{aa}}{2}\right) 
\frac{1}{(2 \pi)^{20}} \int d^4k' d^4q_1 \; d^4q_2 \; d^4q_3 \; d^4q_4 \nonumber\\
&&~~~~~~\cdot
\left[i\frac{\not{q_1} + m}{q_1^2-m^2}\right]\gamma_\tau\frac{\not{q_2} + m}
{q_2^2-m^2}\gamma^\mu \frac{\not{q_3} + m}{q_3^2-m^2}\gamma_\rho \left[i 
\frac{\not{q_4} + m}
{q_4^2-m^2}\right]\left[\frac{-i}{k'^2}\right] \nonumber\\
&&~~~~~~\cdot\left[\int d^4x' e^{ix'\cdot(q_4-p_1)}\right]
\left[\int d^4y' e^{iy'\cdot(p_2-q_1)}\right]
\left[\int d^4y e^{iy\cdot(q_2-q_3+k')}\right]\nonumber\\&&~~~~~~\cdot
\int d^4x \; d^4z \; e^{ix\cdot(q_3-q_4)}
e^{iz\cdot(q_1-q_2)}
\left[C(x-z)^\tau(x-z)^\rho+g^{\tau\rho}E(x-z)^2\right] \nonumber\\
&&~~~=\frac{i^2}{8}(-ieQ)(-ig_s)^2\left(\frac{4}{3}\right)
\frac{1}{(2 \pi)^{8}} 
\left[i\frac{\not{p_2} + m}{p_2^2-m^2}\right]
\int d^4q_2 \; d^4q_3 
\nonumber\\&&~~~~~~\cdot 
\gamma_\tau \left\{
\frac{\not{q_2} + m}{q_2^2-m^2} \gamma^\mu\frac{\not{q_3} + m}{q_3^2-m^2}
\right\}\gamma_\rho
\left[ \frac{-i}{(q_3-q_2)^2}\right]
\int d^4x \; d^4z \; e^{ix\cdot(q_3-p_1)} \cdot
\nonumber\\&&~~~~~~ \cdot e^{iz\cdot(p_2-q_2)}
\left[C(x-z)^\tau(x-z)^\rho+g^{\tau\rho}E(x-z)^2\right]
\left[i\frac{\not{p_1} + m}{p_1^2-m^2}\right]\nonumber\\&&~~~\equiv
\frac{i^2}{8}(-ieQ)(-ig_s)^2\left(\frac{4}{3}\right)
\frac{1}{(2 \pi)^{8}} 
\left[i\frac{\not{p_2} + m}{p_2^2-m^2}\right] I^\mu(p_1,p_2)
\left[i\frac{\not{p_1} + m}{p_1^2-m^2}\right].
\end{eqnarray}

To evaluate $I^\mu$, as defined in the final line of (\ref{qcdGreen}), we make
the following change of variables,
\begin{eqnarray}
{\cal X}\equiv z-x ; {\cal Z}\equiv x+z \nonumber\\
x=\frac{1}{2}({\cal Z}-{\cal X}) ; z=\frac{1}{2}({\cal X}+{\cal Z})
\end{eqnarray}
and we define
\begin{equation}
H^\mu(q_2,q_3) \equiv \left\{
\frac{\not{q_2} + m}{q_2^2-m^2} \gamma^\mu\frac{\not{q_3} + m}{q_3^2-m^2}
\right\}
\end{equation}
\begin{equation}
{\cal P}\equiv \frac{1}{2}(p_1+p_2); \; {\cal Q}\equiv\frac{1}{2}(p_2-p_1)
\end{equation}
We then find that
\begin{eqnarray}\label{Imu}
I^\mu(p_1,p_2) &\equiv&\int d^4q_2 \; d^4q_3 \; \gamma_\tau H^\mu(q_2,q_3)\gamma_\rho
\left[ \frac{-i}{(q_3-q_2)^2}\right]\int d^4x
\; d^4z \; e^{ix\cdot(q_3-p_1)}e^{iz\cdot(p_2-q_2)}  
\nonumber\\&~&\cdot\left[C(x-z)^\tau(x-z)^\rho +g^{\tau\rho}E(x-z)^2\right] 
\nonumber\\&=&
(\frac{1}{2})^4\int d^4q_2 \; d^4q_3 \; \gamma_\tau H^\mu(q_2,q_3)\gamma_\rho
\left[ \frac{-i}{(q_3-q_2)^2}\right]\int d^4{\cal X} d^4{\cal Z}
e^{i{\cal X}\cdot[{\cal P}-\frac{1}{2}(q_2+q_3)]}
\nonumber\\&~&\cdot e^{i{\cal Z}\cdot[{\cal Q}-\frac{1}{2}(q_2-q_3)]}
\left[C {\cal X}^\tau {\cal X}^\rho +
g^{\tau\rho}E {\cal X}^2\right]
\nonumber\\&=&
(\frac{1}{2})^4\int d^4q_2 \; d^4q_3 \; \gamma_\tau H^\mu(q_2,q_3)\gamma_\rho
\left[ \frac{-i}{(q_3-q_2)^2}\right]
\left\{Cg^{\alpha\rho}g^{\beta\tau}+Eg^{\rho\tau}g^{\alpha\beta}\right\}
\nonumber\\&~&\cdot(2\pi)^4\delta^4\left[{\cal Q}-\frac{1}{2}(q_2-q_3)\right] 
\int d^4{\cal X} {\cal X}_\alpha {\cal X}_\beta
e^{i{\cal X}\cdot[{\cal P}-\frac{1}{2}(q_2+q_3)]}
\nonumber\\&=& 
(\frac{1}{2})^4\int d^4q_2 \; d^4q_3 \;\gamma_\tau H^\mu(q_2,q_3)\gamma_\rho
\left[ \frac{-i}{(q_3-q_2)^2}\right]
\left\{Cg^{\alpha\rho}g^{\beta\tau}+Eg^{\rho\tau}g^{\alpha\beta}\right\}
\nonumber\\&~&\cdot(2\pi)^4\delta^4\left[{\cal Q}-\frac{1}{2}(q_2-q_3)\right] 
\left(-\frac{\partial^2}{\partial {\cal P}^\alpha \partial {\cal P}^\beta}
\right)
(2\pi)^4\delta^4\left[{\cal P}-\frac{1}{2}(q_2+q_3)\right]
\nonumber\\&=&  
(\frac{1}{2})^4(2\pi)^8
\left\{Cg^{\alpha\rho}g^{\beta\tau}+Eg^{\rho\tau}g^{\alpha\beta}\right\}
\left(-\frac{\partial^2}{\partial {\cal P}^\alpha \partial {\cal P}^\beta}
\right)\gamma_\tau
H_\sigma\left({\cal P}+{\cal Q},
{\cal P}-{\cal Q}\right)\nonumber\\&~&\cdot 
\gamma_\rho \left[ \frac{-ig^{\mu\sigma}}{4{\cal Q}^2}\right].
\end{eqnarray}
Factorization of the external photon in the final line of (\ref{Imu}), in
conjunction with factorization of the external fermions in the final line of
(\ref{qcdGreen}), allows a clear determination of the gluon-condensate
contribution to the truncated vertex function:
\begin{eqnarray} \label{qcdVertex}
\Delta \Gamma^\sigma(p_2,p_1) &=& 
\frac{i^2}{8}(-ig_s)^2\left(\frac{4}{3}\right)(\frac{1}{2})^4
\frac{1}{4}
\left[Cg^{\alpha\rho}g^{\beta\tau}+Eg^{\rho\tau}g^{\alpha\beta}\right]
\gamma_\tau
\left(-\frac{\partial^2}{\partial {\cal P}^\alpha \partial {\cal P}^\beta}
\right) \nonumber\\&~& \cdot
\left\{ 
\frac{\not{{\cal P}}+\not{{\cal Q}} + m}{({\cal P}+{\cal Q})^2-m^2}
\gamma^\sigma\frac{\not{{\cal P}}-\not{{\cal Q}} + m}{({\cal P}-{\cal Q})^2-m^2}
\right\} \gamma_\rho.
\end{eqnarray}
Define
\begin{equation}
D_1 \equiv ({\cal P}-{\cal Q})^2-m^2 = p_1^2-m^2 ~;~
D_2 \equiv ({\cal P}+{\cal Q})^2-m^2 = p_2^2-m^2.
\end{equation}
The partial derivatives within (\ref{qcdVertex}) are then found to be
\begin{eqnarray} \label{partialH}
&&\frac{\partial^2}{\partial {\cal P}^\alpha \partial {\cal P}^\beta}
\left\{ 
\frac{1}{D_1 D_2} (\not{{\cal P}}+\not{{\cal Q}} + m)
\gamma^\sigma(\not{{\cal P}}-\not{{\cal Q}} + m)\right\}
 \nonumber\\&&~~~=
\left[B_{\alpha\beta}(\not{p_2} + m)
\gamma^\sigma(\not{p_1} + m) + \frac{1}{D_1 D_2}
(\gamma_\beta \gamma^\sigma \gamma_\alpha+
\gamma_\alpha \gamma^\sigma \gamma_\beta )\right] ,
\end{eqnarray}
where
\renewcommand{\theequation}{4.14\alph{equation}}
\setcounter{equation}{0}
\begin{eqnarray} \label{Aalpha}
A^\alpha &\equiv& \frac{\partial}{\partial P_\alpha} \frac{1}{D_1 D_2}
= -\frac{2}{D_1^2 D_2^2}(p_1^\alpha D_2 + p_2^\alpha D_1) \\
B^{\alpha \beta} &\equiv&  \label{BalphaBeta}
\frac{\partial}{\partial P_\alpha}A^\beta = 
2D_1 D_2 A^\alpha A^\beta - \frac{2}{D_1^2 D_2^2}\left[
g^{\alpha\beta}(D_1+D_2)+2(p_1^\alpha p_2^\beta + p_2^\alpha p_1^\beta)\right]
\end{eqnarray}
\renewcommand{\theequation}{4.\arabic{equation}}
\setcounter{equation}{14}
Upon substitution of (\ref{partialH}) into (\ref{qcdVertex}), we find that
\newpage
\begin{eqnarray} \label{qcdVertex2}
\Delta \Gamma^\sigma(p_2,p_1) &=& 
\frac{-g_s^2}{3\cdot 2^7}
\left[Cg^{\alpha\rho}g^{\beta\tau}+Eg^{\rho\tau}g^{\alpha\beta}\right]
\nonumber\\&~&\cdot
\gamma_\tau\left[B_{\alpha\beta}(\not{p_2} + m)
\gamma^\sigma(\not{p_1} + m) + \frac{1}{D_1 D_2}
(\gamma_\beta \gamma^\sigma \gamma_\alpha+
\gamma_\alpha \gamma^\sigma \gamma_\beta )\right]\gamma_\rho
\nonumber\\&=& \frac{-g_s^2}{3\cdot 2^7}
\left\{C\left[ B_{\alpha\beta}\gamma^\beta(\not{p_2} + m)
\gamma^\sigma(\not{p_1} + m)\gamma^\alpha
\right.\right.\nonumber\\&~&~~~~~~~~~~~~~~ \left.
+\frac{1}{D_1 D_2}
(\gamma^\beta \gamma_\beta \gamma^\sigma \gamma_\alpha \gamma^\alpha
+\gamma^\beta \gamma_\alpha \gamma^\sigma \gamma_\beta \gamma^\alpha)\right]+
\nonumber\\&~&\left.~~~~~~~~~~E\left[g^{\alpha \beta}
B_{\alpha\beta}\gamma_\tau(\not{p_2} + m)
\gamma^\sigma(\not{p_1} + m)\gamma^\tau \right.\right.
\nonumber\\&~&\left. \left. ~~~~~~~~~~~~~~+\frac{1}{D_1 D_2}
(\gamma_\tau \gamma^\alpha \gamma^\sigma \gamma_\alpha \gamma^\tau
+\gamma_\tau \gamma_\alpha \gamma^\sigma \gamma^\alpha \gamma^\tau)
\right]\right\}
\end{eqnarray}
We can then sandwich $\Delta \Gamma$ between on-shell spinors,
\begin{eqnarray}
\overline{u}(p_2)(\not{p_2} - m) &=& 0 ,\nonumber\\
(\not{p_1} - m)u(p_1) &=& 0 ,
\end{eqnarray}
and utilize the relations
\begin{eqnarray}
\frac{(\not{p_1} - m)}{p_1^2-m^2}u(p_1) &=& \frac{1}{2m}u(p_1) \nonumber\\
\overline{u}(p_2)\frac{(\not{p_2} - m)}{p_2^2-m^2} &=&
\overline{u}(p_2)  \frac{1}{2m} 
\end{eqnarray}
Care must be taken not to cancel terms in the numerator resulting from naively
applying the on-shell relations without regard for vanishing denominator factors
of $D_1$ and $D_2$ in (\ref{qcdVertex2}).  After a fair amount of Dirac algebra,
the cautious application of these identities [without yet setting
$(p_1^2=p_2^2=m^2)$] to (\ref{qcdVertex2}) taken between on-shell spinors yields
the following result for the gluon-condensate contribution to the truncated
electromagnetic vertex function [$D_1=D_2 \equiv {\cal D}$]:
\newpage
\begin{eqnarray} \label{gluonVertex}
&&\overline{u}(p_2)\Delta \Gamma^\sigma (p_2,p_1)u(p_1) \nonumber\\&&~=
\frac{-g_s^2}{3\cdot2^7}\overline{u}(p_2)\left\{\left[
\frac{20C+8E}{{\cal D}^2}+\frac{8}{{\cal D}^3}\left(\frac{2E(2m^2+p_1 \cdot
p_2)}{{\cal D}}-C \right)\left(2p_1 \cdot p_2+{\cal D} + \frac{{\cal D}^2}{4m^2}
\right)
\right.\right. \nonumber\\&&~~~~~~\left.
+\frac{8C}{{\cal D}^4}\left(2m^4+2(p_1 \cdot p_2)^2+
\left(4m^2+7{\cal D}+\frac{{\cal D}^2}{4m^2}\right) p_1 \cdot p_2 + 3{\cal D}^2
+4m^2{\cal D}\right)\right]\gamma^\sigma 
\nonumber\\&&~~~~~~
+(p_1^\sigma +p_2^\sigma)\left[-\frac{8}{m{\cal D}^2}\left(
\frac{2E(2m^2+p_1 \cdot p_2)}{{\cal D}}-C\right)
-\frac{8C}{{\cal D}^3}\left(\frac{3{\cal D}}{2m}+2m\right)
\right. \nonumber\\&&~~~~~~\left.\left. 
- \frac{4C}{{\cal D}^3}\left(\frac{2p_1 \cdot p_2}{m}-\frac{{\cal D}}{2m}
\right)\right]\right\}u(p_1)
\end{eqnarray}
We now utilize the decomposition (\ref{qedVertexStructure}) to define the 
gluon-condensate contributions to $R(q^2)$ and $S(q^2)$ at $q^2 \rightarrow 0$.  Since
${\cal D} \rightarrow 0$ on-shell, such contributions are highly divergent:
\begin{equation}\label{R0}
[R(0)]_{<G^2>} =  \frac{-g_s^2}{3 \cdot 2^7 e^2 Q^2}
\left[\frac{(96E+64C)m^4}{{\cal D}^4} + {\cal O}\left(\frac{m^2}{{\cal D}^3}
\right) \right]
\end{equation}
\begin{equation}\label{S0}
[S(0)]_{<G^2>} =  \frac{-g_s^2}{3 \cdot 2^7 e^2 Q^2}
\left[-\frac{(24E+12C)m^2}{{\cal D}^3} - \frac{C}{{\cal D}^2} \right]
\end{equation}
It is evident from substitution of (\ref{S0}) into (\ref{amm}) that the 
gluon-condensate contribution to the anomalous magnetic moment of a quark appears to
be divergent, a consequence of the fact that $S(0)$ is now divergent, as
opposed to the purely perturbative case (Section \ref{review}) where $S(0)$ is 
manifestly finite.  Moreover, the
problem is not remedied by any cancellation between $<G^2>$ and $<(\partial_\mu
A^\mu)^2>$ terms contributing to $C$ and $E$ [e.g. (4.4)], since no such
cancellation can simultaneously render $(24E+12C)=0$ to remove the ${\cal
D}^{-3}$ divergence and $C=0$ to remove the ${\cal D}^{-2}$ divergence.

We note, of course, that finite quark magnetic moments appear to characterize
constituent quarks in a number of phenomenologically successful applications, as
discussed in Section \ref{introduction}.  In Section \ref{discussion}, we
will explore whether (\ref{amm}) is indeed applicable to the anomalous magnetic
moment of a condensing fermion.
\newpage
\section{Discussion} \label{discussion}
\renewcommand{\theequation}{5.\arabic{equation}}
\setcounter{equation}{0}

\subsection{{\it Quark Condensate Contributions}}
Let us begin by considering {\it in isolation} the fermion-antifermion
condensate contribution to $S(q^2)$.  If we disregard for now the nonzero
imaginary part (\ref{imS}), a dispersive contribution necessarily indicative of
physical intermediate states, we see from (\ref{reS}) that the
fermion-antifermion condensate contribution to ${\cal K} F_2(0)$ is zero, as the
real part of $S(0)$ is zero.  In and of itself, this result decouples any
quark-antiquark condensate effects from the quark's anomalous magnetic moment. 
As discussed in Section \ref{introduction}, however, purely perturbative effects are
expected to give a substantial contribution to the quark's anomalous magnetic
moment that would create a discrepancy between constituent-quark masses
characterizing baryon spectroscopy and those characterizing baryon magnetic
moments.  The absence of any offsetting condensate contribution, however
interesting in itself, leaves us with these phenomenological issues unresolved.

The occurence in $S(q^2)$ of a nonzero imaginary part when $0<q^2<4m^2$ is also
of interest.  The purely perturbative contribution to $S(q^2)$ acquires an
imaginary part only when $q^2>4m^2$, corresponding to being above 
the kinematic threshold
for the production of a physical fermion-antifermion pair.  In other words, the
Feynman amplitude for the vertex correction acquires a dispersive component
associated with the presence of a physical fermion-antifermion intermediate
state.  In the presence of a vacuum $|\Omega>$ that permits $<f \bar{f}>$
condensation ({\it i.e.,} the QCD vacuum), the kinematic threshold for this
dispersive component is lowered from $q^2=4m^2$ to $q^2=0$.

To understand why this occurs, it may be useful to recapitulate the input
assumptions of the calculation presented in Section \ref{fermContrib}.  A single
dynamical mass is assumed to characterize both the perturbative
(\ref{fermionProp}) and
the nonperturbative (\ref{F}) fermion propagator, the latter being a reflection
of the chiral noninvariance of the vacuum $|\Omega>$.  Any attempt to
differentiate between perturbative and nonperturbative propagator masses
compromises the gauge invariance of ${\cal K} F_2(0)$, unless the Feynman rules
utilized in obtaining $S(q^2)$ are themselves modified by 1PI contributions
necessary to retain consistency with BRST identities.$^{\cite{ahm91}}$  We have
chosen to utilize a single dynamical mass to characterize the Feynman amplitude
in order to avoid these complications.  Indeed, a similar need to equilibrate
nonperturbative and perturbative propagator masses in order to maintain gauge
invariance has been demonstrated within the context of condensate contributions
to two-point functions,$^{\cite{ahm89}}$ and such two-point functions are also seen
to acquire dispersive contributions for values of momentum-squared between zero
and $4m^2$.$^{\cite{elias93}}$  In any case, the mass we are using does not arise
because of a lagrangian mass term, but rather as a consequence of the chiral
nonivariance of the vacuum $|\Omega>$.  Such violation of explicit chiral
symmetry necessarily entails the presence of a massless Goldstone boson in the
particle spectrum, in which case the extension of the dispersive component's
kinematical domain from $q^2>4m^2$ to $q^2>0$ must be understood as
corresponding to the production of "physical" zero-mass Goldstone bosons. 
Within a QCD context, such Goldstone bosons are, of course, pions.  We see,
therefore, that the lowering of the threshold for $S(q^2)$'s dispersive
component from $4m^2$ to zero may be a direct indication of QCD's transition
from a gauge theory of quarks and gluons to a theory of low-energy hadron
physics.

\subsection{{\it Gluon-Condensate Contributions}}

We have seen in Section \ref{gluContrib} that the gluon-condensate contributions
to $S(q^2)$ are divergent on-shell, leading via (\ref{amm}) to a divergent
anomalous magnetic moment.  It is worth examining whether the set of assumptions
leading to (\ref{amm}) remains applicable when such nonperturbative contributions
are present.

We first note that on the ${\cal D} =0$ mass shell, the gluon-condensate
contributions to $R(q^2)$ are even more divergent than the gluon-condensate
contributions to $S(q^2)$; the former diverge like $1/ {\cal D}^4$ (\ref{R0})
whereas the latter diverge like $1/{\cal D}^3$ (\ref{S0}).  To leading order in
$\alpha_s$, there also exist gluon-condensate contributions to $R(q^2)$ arising
from self-energy insertions (Fig. 5).  However, gluon-condensate contributions
to self-energies$^{\cite{bagan94}}$
\begin{equation}
\Sigma(p)=\frac{\pi m^3 \alpha_s}{3(p^2-m^2)} <\Omega|:(\partial \cdot
A)^2:|\Omega> + \frac{\pi \alpha_s [(p^2-3m^2) \not{p} +
3m^3]}{9(p^2-m^2)^3}<G^2>~,
\end{equation}
when inserted into the Fig. 5 graphs, fail to cancel the leading $1/{\cal D}^4$
divergence of $R(q^2)$ arising from Fig. 4.

We note that QCD vacuum condensates are supposed to be renormalization-group
(RG) invariant structures.  Indeed, the RG-invariant dimension-4
gluon-condensate is not $<G^2>$, but rather is $<\beta(\alpha_s)G^2/\alpha_s>$.  To
leading order in $\alpha_s$, we note that $\beta(\alpha_s)/\alpha_s$ is
proportional to $\alpha_s$ itself, in which case the RG-invariant condensate
absorbs two factors of the strong coupling $g_s$; {\it i.e.,} if (\ref{R0})
and (\ref{S0}) are to be expressed in terms of RG invariants, then
\begin{equation} \label{G2}
g_s^2 <G^2> \rightarrow 4 \pi <\alpha_s G^2>~.
\end{equation}
The condensate $<\alpha_sG^2>$ is itself familiar to QCD sum-rule
applications,$^{\cite{shifman}}$ and has a sum-rule estimated magnitude ($0.045
\; GeV^4$) somewhat larger than $(\Lambda_{QCD})^4$, indicative of a
dimensionally-appropriate QCD
momentum-scaling that is not subject to additional perturbative suppression.
This suggests that (\ref{R0}) and (\ref{S0}) are in fact {\it order-unity} in
the perturbation series.  If one considers purely perturbative and
gluon-condensate contributions together with other (perturbatively suppressed)
contributions, one will then find for the truncated
vertex $\bar{u}(p_2)[\gamma^\tau + \Gamma^\tau(p_2,p_1)]u(p_1)$, analogous to
(\ref{qedVertexStructure}), that
\begin{eqnarray}
\label{5:gamma}
\Gamma^\tau(p_2,p_1)=\gamma^\tau [ [R(q^2)]_{<G^2>} + {\cal O}(g_s^2,e^2)] +
\frac{2(p_1^\tau+p_2^\tau)}{m}[[S(q^2)]_{<G^2>} + {\cal O}(g_s^2,e^2)]
\end{eqnarray}
where, from (\ref{R0},\ref{S0}) and (\ref{G2}),
\renewcommand{\theequation} {5.4.\alph{equation}}
\setcounter{equation}{0}
\begin{eqnarray}
[R(0)]_{<G^2>} &\sim& <\alpha_s G^2>m^4/{\cal D}^4~,\\
\renewcommand{\theequation} {5.4b}
[S(0)]_{<G^2>} &\sim& <\alpha_s G^2>m^2/{\cal D}^3~.
\end{eqnarray}
\renewcommand{\theequation}{5.\arabic{equation}}
\setcounter{equation}{4}
Eq. (\ref{5:gamma}) differs from (\ref{qedVertexStructure}), the definition of $R(q^2)$ and
$S(q^2)$ for the purely perturbative case, only by omission of an overall factor
of $e^2Q^2$, indicative that the leading vertex correction is no longer one
order of perturbation theory removed from unity.  As a consequence of this
change, we now identify (\ref{5:gamma}) with the final line of (\ref{211}).  Upon
application of the renormalization condition requiring $F_1(0)$ to be
constrained to unity, Eq. (\ref{amm}) is then replaced with the following
relation:
\begin{eqnarray} \label{kappaEquals0}
{\cal K} F_2(0) &=& 
\lim_{q^2 \rightarrow 0} \frac{-4S(q^2)}{1+R(q^2)+4S(q^2)} \nonumber\\&&~
\nonumber\\
&=& \lim_{{\cal D} \rightarrow 0} \frac{-4[S]_{<G^2>}+{\cal O}(g_s^2,e^2)}{1+[R]_{<G^2>}+4[S]_{<G^2>}
+{\cal O}(g_s^2,e^2)} \nonumber\\&&~\nonumber\\
&=& 0 .
\end{eqnarray}
The last line follows trivially from (5.4).  It should be noted that
(\ref{kappaEquals0}) is not just a statement about the gluon-condensate
contribution, but rather a statement of what happens to all contributions to the
anomalous magnetic moment [{\it i.e.,} the ${\cal O}(g_s^2,e^2)$ contributions]
in the presence of a gluon-condensate contribution that is not suppressed
perturbatively.  All other condensate contributions, like the quark condensate
contribution of Section \ref{fermContrib}, are expected to be suppressed by at
least one factor of $g_s^2$.

Thus, the presence of a QCD vacuum $|\Omega>$ that permits gluon condensation
appears to preclude the possibility of quarks acquiring any anomalous magnetic
moment at all.  This would suggest that quarks behave like naive Dirac fermions
with magneton $eQ/2m$, where $m$ is of order $350\;MeV$ [recall that $m$ has
already been identified as a dynamical mass].  The idea that constituent quarks
act like fundamental Dirac fermions is supported in other contexts as
well.$^{\cite{wein}}$  Indeed, the absence of an anomalous magnetic moment for
constituent quarks is of some phenomenological utility, in that it obviates the
discrepancy anticipated on purely perturbative grounds (Section
\ref{introduction}) between constituent quark masses characterizing baryon
spectroscopy and those characterizing baryon magnetic moments.

Of course, there are necessarily {\it caveats} to all this -- salient among
these is the use of a strong coupling at $q^2=0$, as discussed in Section
\ref{introduction}.  The use of condensates obtained in a covariant gauge is also a
nonstandard technique which, in principle, includes the presence of a gauge
noninvariant condensate $<(\partial \cdot A)^2>$.  This quantity can be shown to
cancel algebraically from some other physical amplitudes,$^{\cite{bagan94}}$ and it is
somewhat disconcerting to have it not do so here [except in the sense that 
$<(\partial \cdot A)^2>$, along with both $<\alpha_s G^2>$ and $<q \bar{q}>$,
appears
ultimately to be decoupled from ${\cal K} F_2(0)$].  However, we believe there 
is at least
methodological merit in the field theoretical calculation of condensate
contributions to a vertex function.  The techniques now exist to do such
calculations; the challenge remaining is to make sense of the results.
\bigskip

\noindent {\bf Acknowledgements}
\bigskip

\noindent
Kurt Haller's continuing body of work addressing the fundamental underpinnings
of quantum electrodynamics has provided both inspiration and a model for
formulating and grappling with unresolved issues within quantum field theory. 
V.E. is particularly grateful to Kurt Haller for a number of seminal discussions
on the field-theoretical consistency of vacuum condensates within abelian and
nonabelian quantum field theories.  The authors also acknowledge research
support from the Natural Sciences and Engineering Research Council of Canada.

\end{document}